\documentclass[showpacs, oneside, twocolumn, amsmath, amssymb, prd, nofootinbib, superscriptaddress]{revtex4-1}

\usepackage{ulem}

\usepackage{cases}
\usepackage{amsmath}
\usepackage{amssymb}
\usepackage{amsfonts}
\usepackage{amssymb}
\usepackage{dcolumn}
\usepackage{bm}
\usepackage{bbm}
\usepackage{graphicx}
\usepackage{xcolor}
\usepackage{array}
\usepackage{subfigure}
\usepackage{hyperref}
\usepackage{wasysym}

\newcommand{\be}{\begin{equation}}
\newcommand{\ee}{\end{equation}}
\newcommand{\ba}{\begin{eqnarray}}
\newcommand{\ea}{\end{eqnarray}}

\newcommand{\gsim}{\mathrel{\hbox{\rlap{\lower.55ex \hbox {$\sim$}}
                   \kern-.3em \raise.4ex \hbox{$>$}}}}
\newcommand{\lsim}{\mathrel{\hbox{\rlap{\lower.55ex \hbox {$\sim$}}
                   \kern-.3em \raise.4ex \hbox{$<$}}}}

\hypersetup{colorlinks=true,
            breaklinks=true,
            pdfstartview=Fit,
            linkcolor=blue,
            citecolor=blue,
            urlcolor=blue}

\newcommand\R{\mathcal{R}}
\newcommand\s{\mathcal{S}}

\begin{document}
\title{Shift-Symmetric Orbital Inflation: single field or multi-field?}

\author{Ana Ach{\'u}carro}
\email{achucar@lorentz.leidenuniv.nl}
\affiliation{Lorentz Institute for Theoretical Physics, Leiden University, 2333 CA Leiden, The Netherlands}
\affiliation{Department of Theoretical Physics, University of the Basque Country, 48080 Bilbao, Spain}

\author{Edmund J. Copeland}%
 \email{ed.copeland@nottingham.ac.uk}
\affiliation{School of Physics and Astronomy, University of Nottingham, Nottingham NG7 2RD, UK\\
}%

\author{Oksana Iarygina}%
 \email{iarygina@lorentz.leidenuniv.nl}
\affiliation{Lorentz Institute for Theoretical Physics, Leiden University, 2333 CA Leiden, The Netherlands}%

\author{Gonzalo A. Palma}
 \email{gpalmaquilod@ing.uchile.cl}
 \affiliation{Grupo de Cosmolog\'ia y Astrof\'isica Te\'orica, Departamento de F\'{i}sica, FCFM,\\ \mbox{Universidad de Chile}, Blanco Encalada 2008, Santiago, Chile}

\author{Dong-Gang Wang}
\email{wdgang@strw.leidenuniv.nl}
\affiliation{Lorentz Institute for Theoretical Physics, Leiden University, 2333 CA Leiden, The Netherlands}
\affiliation{Leiden Observatory, Leiden University, 2300 RA Leiden, The Netherlands}

\author{Yvette Welling}
\email{welling@strw.leidenuniv.nl}
\affiliation{Lorentz Institute for Theoretical Physics, Leiden University, 2333 CA Leiden, The Netherlands}
\affiliation{Leiden Observatory, Leiden University, 2300 RA Leiden, The Netherlands}
\affiliation{Deutsches Elektronen-Synchrotron DESY, Notkestra{\ss}e 85, 22607 Hamburg, Germany}

\date{\today}

\begin{abstract}
We present a new class of two-field inflationary attractor models, known as {\it `shift-symmetric orbital inflation'},
whose behaviour is strongly multi-field but whose predictions are remarkably close to those of single-field inflation.
In these models, the field space metric and potential are such that the inflaton trajectory is along an `angular' isometry direction  whose `radius' is constant but {\it arbitrary}. As a result, the radial (isocurvature) perturbations away from the trajectory are exactly {\it massless} and they freeze on superhorizon scales. These models are the first exact realization of the `ultra-light isocurvature' scenario, previously described in the literature, where a combined shift symmetry emerges between the curvature and isocurvature perturbations and results in primordial perturbation spectra that are entirely consistent with current observations. Due to the turning trajectory,  the radial perturbation sources the tangential (curvature) perturbation and makes it grow linearly in time.
As a result, only one degree of freedom ({\it i.e.} the one from isocurvature modes) is responsible for the primordial observables at the end of  inflation, which yields the same phenomenology as in single-field inflation. In particular, isocurvature perturbations and local non-Gaussianity are highly suppressed here, even if the inflationary dynamics is truly multi-field. We comment on the  generalization to models with more than two fields.

\end{abstract}

\maketitle

\section{Introduction}

{Single field slow roll inflation is the leading explanation for the observations through the CMB \cite{Akrami:2018odb} that primordial perturbations are very close to Gaussian and adiabatic, yet embedding it in an ultraviolet complete theory such as string theory is notoriously difficult. Moduli fields arising from string compactifications require stabilizing to realize single field inflation \cite{Baumann:2014nda}, and large field excursions test the validity of using four dimensional effective theories}\footnote{{The recent swampland debate highlights the importance of finding viable scenarios for inflation that are not strictly single-field. See, for instance, the discussion in} \cite{Achucarro:2018vey} {as compared to} \cite{Obied:2018sgi, Agrawal:2018own}}.

In the usual understanding, light fields during inflation may lead to isocurvature perturbations and local non-Gaussianity tightly constrained by current observations.
However, it has been suggested recently that inflation with non-stabilized light fields on an axion-dilaton system can be compatible with the latest CMB data \cite{Kobayashi:2010fm, Turzynski:2014tza, Cremonini:2010sv, Cremonini:2010ua, vandeBruck:2014ata, Achucarro:2016fby, Achucarro:2017ing}.
In particular, it was pointed out in \cite{Achucarro:2016fby} that, when the perturbations orthogonal to the trajectory are {\it massless} but efficiently {\it coupled} to the inflaton, the isocurvature modes are dynamically suppressed\footnote{Observational constraints on isocurvature perturbations do not directly constrain the generation of primordial isocurvature fluctuations during inflation. The existence of isocurvature perturbations in the CMB depends on how inflationary isocurvature fluctuations decayed during reheating, hence, while inflationary isocurvature perturbations are necessary for the existence of isocurvature perturbations in the CMB, the absence of the latter cannot be used to rule out multi-field inflation.}. This is the ``ultra-light isocurvature" scenario.

In this paper we provide for the first time a family of exact models of inflation in which the multi-field effects are significant, but the phenomenology remains similar to single field inflation.
The models combine two ingredients: First, the inflaton trajectory proceeds along an isometry direction of the field space, so it is Orbital Inflation  in the sense of  \cite{Achucarro:2019mea,Welling:2019bib}. This ensures time independence of the coupling between the radial and tangential inflationary perturbations.  Second, the trajectory can have an {\it arbitrary} radius (within some range described below), and a constant radius is proven to be a neutrally stable attractor (see Appendix B). Hence, isocurvature perturbations become exactly massless. The two ingredients, combined, guarantee that the sourcing of the curvature perturbation is sustained over many e-folds of inflationary expansion. The action for the perturbations inherits a symmetry between background solutions that is not manifest in the potential or in the Lagrangian. We show that, at the end of inflation, only the isocurvature degree of freedom is responsible for the generation of primordial observables, but perturbations still remain adiabatic and Gaussian. We call this scenario {\it shift-symmetric orbital inflation}.

Crucially this scenario provides a new direction to explore inflation and a potential resolution to some of the problems faced by the embedding of inflation in string theory.
That is, in the construction of inflationary models wherein every modulus is stabilized except for the inflaton, one could be missing less restrictive realizations of inflation compatible with current observational constraints. We set $\hbar = c = 1$ and the reduced Planck mass $M_p\equiv (8\pi G)^{-1/2}=1$, where $G$ is Newton's contant.

\section{A toy model}

\begin{figure}[h]
\includegraphics[width=0.4\textwidth]{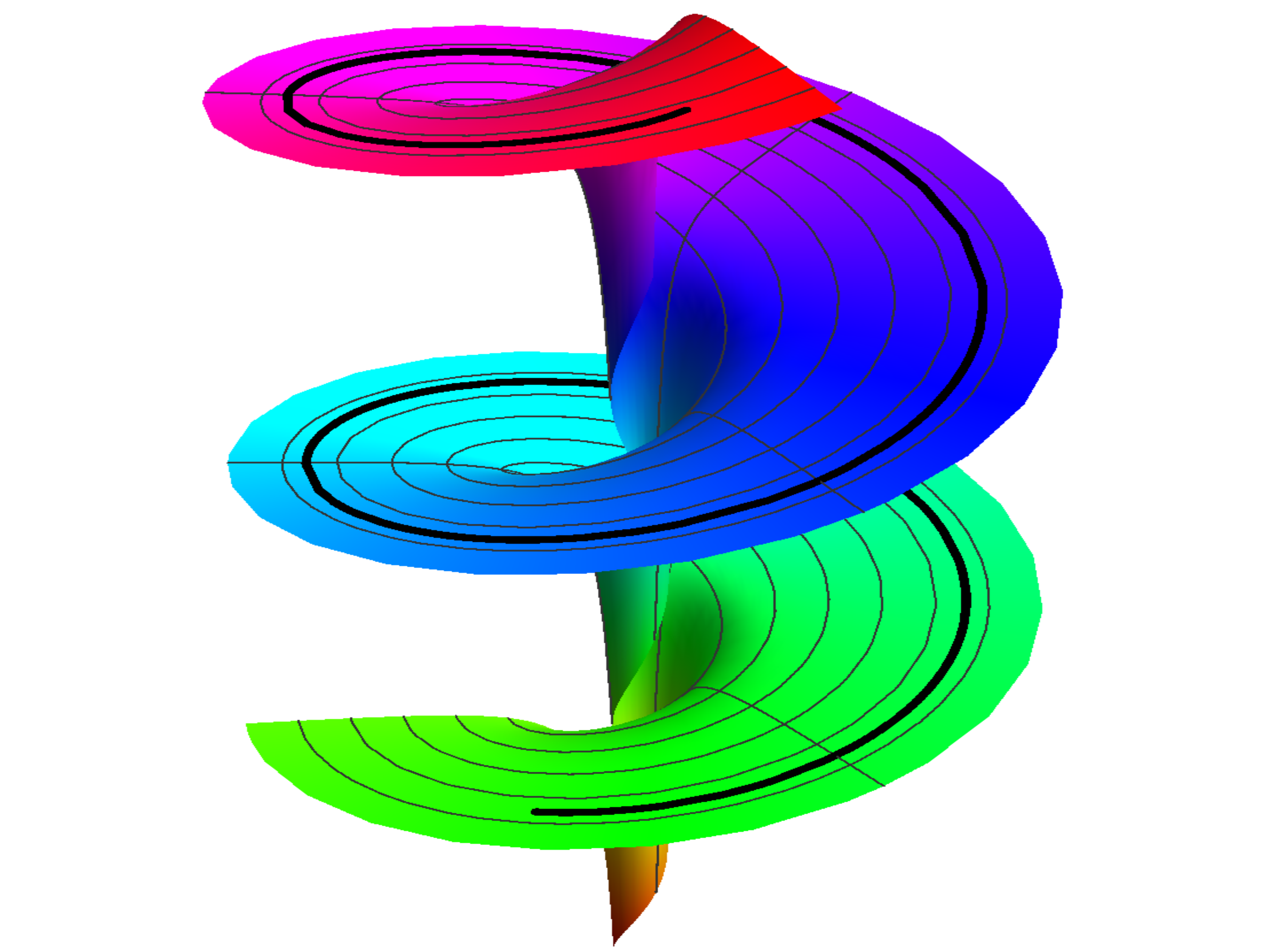}
\centering
\caption{The toy model potential $V(\rho, \theta)$ given in \eqref{eqn:toymodel} together with a typical inflationary trajectory indicated with the solid black line.}
\label{fig:toymodelpotential}
\end{figure}

To illustrate the idea, we first consider the following Lagrangian in flat field space with polar coordinates (illustrated in Fig.~\ref{fig:toymodelpotential})
\begin{equation}
\mathcal{L} = \frac{1}{2}\left[\rho^2(\partial\theta)^2+(\partial\rho)^2\right]
-\frac{1}{2} m^2  \left(\theta^2 - \frac{2}{3\rho^2}\right).
\label{eqn:toymodel}
\end{equation}
The potential has a monodromy in the angular coordinate, and although it is unbounded at $\rho\rightarrow0$, inflation only takes place in the physically consistent regime where $V(\rho, \theta)>0$. Moreover, as shown in the perturbation analysis below, our study is restricted to radii that cannot be too small. Therefore, we only care about the local form of the potential close to the inflationary trajectory, which we assume is captured well by~\eqref{eqn:toymodel}. 
In general, it is difficult to solve the background equations analytically in such a system.
However, this model has the following exact {neutrally} stable solutions at any radius {(see Fig. \ref{fig:toymodelpotential})}
\begin{equation}\label{eqn:exactsolutionflatfieldmetric}
 \rho = \rho_0 , \quad \dot\theta = \pm \sqrt{\frac{2}{3}}\frac{m }{\rho_0^2} .
\end{equation}
The Friedmann equation becomes $H^2 = {m^2\theta^2}/{6}$ on the attractor, where $H$ is the Hubble parameter, and
the first slow-roll parameter is $\epsilon \equiv -\dot H/H^2 = \frac{2 }{\rho_0^2 \theta^2}$.
This trajectory is nongeodesic in field space, with turning effects that depend on the radius $\kappa$ of the trajectory. Note that here $\kappa = \rho_0$ but, if the field space geometry is curved, $\kappa$ will be a more general function of $\rho_0$.

The situation is reminiscent of circular orbits in a spherically symmetric gravitational field, where the centripetal force stabilizes the radial direction, and the inflaton can circle at any radius with the corresponding angular velocity.
For the field system on the cosmological background, only the isometric circular orbits appear, and we need to break the shift symmetry of $\theta$ in the potential to overcome the Hubble friction.
We can label each solution by a continuous parameter $c$ with the corresponding map
\begin{equation}
 \rho_c = \rho_0+ c , \quad \left(\theta^2_c\right)^\prime =\frac{\left(\theta^2_0\right)^\prime}{(1+c /\kappa)^2} ,
 \label{eqn:toymodelbackgroundmap}
\end{equation}
where the prime $\prime$ denotes a derivative with respect to efolds $d/dN=d/(Hdt)$.
This transformation identifies all the trajectories in \eqref{eqn:exactsolutionflatfieldmetric} and hints at the existence of a shift symmetry for the perturbations. In flat gauge, the isocurvature perturbation $\sigma$ is associated with $\delta \rho$ and the curvature perturbation $\R$ with $\frac{\rho}{\sqrt{2\epsilon}}  \delta \theta$, which equals $\tfrac{1}{4}\rho^2 \delta\left(\theta^2\right)$ in this toy model. To find the effect of the transformation on the perturbations, we split $\rho = \rho_0 +\sigma $ and $\left(\theta^2\right)^\prime = \left(\theta_0^2\right)^\prime\left(1 - \R^\prime \right)$. This allows us to determine how a small $c$ changes $\sigma$ and $\R^\prime$. In the long wavelength limit every transformed set of perturbations $(\sigma_c, \R^\prime_c)$ provide a new solution to the equations of motion. This is because homogeneous perturbations map background solutions onto each other. Therefore, we expect the following \textit{symmetry} for linearized perturbations
\begin{equation} \label{symmetry}
 \sigma \rightarrow \sigma + c, \quad \R' \rightarrow \R' + \frac{2 }{\kappa}c.
\end{equation}
Given the shift symmetry of $\sigma$, the isocurvature perturbation is expected to be massless and freeze after horizon-exit. Meanwhile, the symmetry also indicates that $\R$ has a growing solution that is dictated by the constant $\sigma$ on superhorizon scales. 

To get an intuitive notion of the perturbations behavior, we employ the $\delta N$ formalism \cite{Starobinsky:1986fxa, Salopek:1990jq, Sasaki:1995aw, Sasaki:1998ug, Lee:2005bb}.
From the Friedmann equation and the exact solution \eqref{eqn:exactsolutionflatfieldmetric}, the number of efolds until the end of inflation is $N = {\rho^2\theta^2}/{4}-1/2$.
The curvature perturbation at the end of inflation is
\begin{equation}
\R(k_\ast)= \delta N \simeq
\frac{1}{\sqrt{2\epsilon_\ast}}(\rho \delta\theta)_\ast+\frac{2N_\ast}{\kappa}\delta\rho_\ast ,
\end{equation}
where $(\rho \delta\theta)_\ast$ and $\delta\rho_\ast$ are field fluctuations with typical amplitude $\frac{H_\ast}{2\pi}$ at horizon-exit of the $k_\ast$ mode.
This yields the following power spectrum of curvature perturbations
\begin{equation}
 P_\R(k_\ast) \simeq \frac{H_\ast^2}{4\pi^2}\left(\frac{1}{2\epsilon_\ast} + \frac{4 N_\ast^2}{\kappa^ 2}\right).
 \label{eqn:toymodelpowerspectrum}
\end{equation}
Here the first contribution has an adiabatic origin, just like in the single-field models, and the second term corresponds to the conversion from isocurvature to curvature modes on superhorizon scales.
When the radius of the trajectory is small enough, namely $8\epsilon_\ast \ll {\kappa^2} \ll 8\epsilon_\ast N_\ast^2 \approx 4N_\ast$,
the second term in \eqref{eqn:toymodelpowerspectrum} dominates. Then the final power spectrum becomes
$ P_\R(k_\ast) \simeq H_\ast^2 N_\ast^2/(\pi^2\kappa^2) $, which is generated by one single degree of freedom -- the isocurvature mode.

\section{Shift-symmetric orbital inflation}

To construct generic models with the above properties,
we begin with an axion-dilaton system in a non-trivial field manifold $(\theta, \rho)$ with kinetic term $K = -\frac{1}{2}\left(f(\rho)\partial_\mu\theta \partial^\mu\theta + \partial_\mu\rho\partial^\mu\rho\right)$.
This field space, of curvature $\mathbb{R}={f_\rho^2}/{2 f^2}-{f_{\rho\rho}}/{f}$, arises generically from UV completions of inflation in quantum gravity or from an effective field theory (EFT) viewpoint.
To realize shift-symmetric orbital inflation, we assume the inflationary trajectory to be isometric, {\it i.e.} along the $\theta$ direction  at \textit{any} (constant) radius in field space.
The potential can be derived by generalizing the Hamilton-Jacobi formalism \cite{Muslimov:1990be, Salopek:1990jq,Lidsey:1991zp,Copeland:1993jj} to a two-field system (See Appendix A).
It has the general form
\begin{equation} \label{potentials}
V= 3H^2 - 2 \frac{H_\theta^2}{f(\rho)} ,
\end{equation}
where $H$ is a function of $\theta$ only, $H_\theta \equiv dH/d\theta$ and $f(\rho)>0$. The model \eqref{eqn:toymodel} is recovered for $H\propto \theta$ and $f(\rho)=\rho^2$,  corresponding to a flat field space parametrized by polar coordinates. This non-linear system admits exact solutions
\be\label{eqn:exactsolutionbgchapexactmodels}
\dot\theta  = -2 \frac{H_\theta}{f}, ~~~~ \rho = \rho_0 .
\ee
Thus the inflaton moves in an orbit of constant radius, as ensured by the Hamilton-Jacobi formalism.
As in the toy model, this trajectory is not along a geodesic.
Here the tangent and normal vectors to the trajectory are $\mathcal{T}^a=1/\sqrt{f}(1,0)$ and $\mathcal{N}^a=(0,1)$, and the radius of the turning trajectory is a constant given by $\kappa = {2 f}/{f_\rho}$.
It follows that all these trajectories are {\it neutrally stable}: a small perturbation orthogonal to a given orbital trajectory will bring us to one of the neighbouring trajectories.
The attractor behaviour is explicitly demonstrated in Appendix B.

\section{Analysis of perturbations}

In flat gauge, the comoving curvature perturbation $\R$ is defined as the projection of the field perturbation along the inflationary trajectory $\R = \frac{1}{\sqrt{2\epsilon}}\mathcal{T}_a \delta\phi^a$, and the isocurvature perturbation $\sigma$ corresponds to the orthogonal projection $\sigma = {\mathcal{N}_a \delta \phi^a}$.
Then for generic multi-field models, the quadratic action of perturbations takes the following form \cite{Achucarro:2016fby}
\begin{equation}\small
 S^{(2)} = \frac{1}{2} \int d^4 x a^3 \left[2\epsilon\left(\dot{\R}-\frac{2H}{\kappa} \sigma\right)^2+\dot{\sigma}^2 -\mu^2 \sigma^2 + \ldots \right] ,
 \label{eqn:quadraticactionperturbations}
\end{equation}
where ellipses stand for the gradient terms $-(\partial_i \sigma)^2 - 2\epsilon (\partial_i \R)^2$.
The interaction between curvature and isocurvature modes is given by the term $a^3({{8\epsilon}H}/{\kappa})\dot\R\sigma$.
To guarantee perturbative analysis we require that $\sqrt{8\epsilon}/\kappa\ll1$~\cite{Achucarro:2016fby, Chen:2018uul}.
The mass of entropy perturbations is defined as
$ \mu^2 \equiv V_{{NN}} + \epsilon H^2\left(\mathbb{R} + {6}/{\kappa^2}\right)$,
where the first term is obtained from the standard Hessian of the potential $ V_{{NN}} \equiv \mathcal{N}^a \mathcal{N}^b \left(V_{ab} - \Gamma^c_{ab} V_c \right)$,
the second and third terms correspond to the field space curvature and turning contributions respectively.

For shift-symmetric orbital inflation, we expect the isocurvature perturbations to be exactly massless, as in the toy model, {and this is confirmed by using  \eqref{eqn:exactsolutionbgchapexactmodels} to show $\mu^2=0$}.
This implies that the quadratic action \eqref{eqn:quadraticactionperturbations} has the combined shift symmetry \eqref{symmetry}, as in the toy model. The power spectra of perturbations in the massless limit can be directly estimated from the coupled evolution of perturbations \cite{Achucarro:2016fby}.
When $\mu = 0$, the linearized system simplifies in the superhorizon limit, yielding
\be
  {\R'_k}=\frac{2 }{\kappa} \sigma_k ,~~~~ \sigma_k = \frac{H_\ast}{2\pi} ,
  \label{eqn:eomsigmash}
\ee
where $*$ denotes evaluation at the time of horizon crossing.
That is, on superhorizon scales the isocurvature perturbation quickly becomes a constant, and it sources the growth of $\R$.
At the end of inflation, the primordial curvature perturbation can be expressed as
$\R_k=\R_\ast +{2N_\ast } \sigma_k/{\kappa}  $,
where the first term is the curvature perturbation amplitude at horizon-exit, and the second term comes from the isocurvature source.
Thus these two contributions are uncorrelated with each other, and
the dimensionless power spectrum for $\R$ is given by
\be \label{spectrum}
 P_\R = \frac{H_*^2}{8\pi^2 \epsilon_\ast} \left(1+   \mathcal{C} \right) ,
\ee
where $\mathcal{C}={8\epsilon_\ast N_\ast^2}/{\kappa^2}$ represents the contribution from isocurvature modes.
This result agrees with the $\delta N$ calculation for the toy model given in \eqref{eqn:toymodelpowerspectrum}.
The full calculation via the in-in formalism gives the same answer up to subleading corrections \cite{Achucarro:2016fby}.
Note that the power spectrum is completely determined by the isocurvature perturbations
if $\mathcal{C}\gg1$, which corresponds to trajectories with a small radius $\kappa$ or, equivalently, significant turning effects {with $8\epsilon_\ast \ll {\kappa^2} \ll 8\epsilon_\ast N_\ast^2$}.
Thus at the end of inflation, curvature perturbations are highly enhanced compared to the ones at horizon-exit.
Meanwhile, the isocurvature power spectrum for $\s\equiv \sigma/\sqrt{2\epsilon}$ remains unchanged as $P_\s=\frac{H_*^2}{8\pi^2 \epsilon_\ast}$.
Therefore, the amplitude of the isocurvature perturbation is dynamically suppressed, {\it i.e.}
$P_\s/P_\R\simeq {1}/{\mathcal{C}}\ll 1$. The details of how $P_\s \neq 0$ can generate isocurvature components in the CMB are rather model-dependent, and one cannot automatically claim that a suppressed ratio $P_\s/P_\R$ is compatible with observations. However, if $\R$ and $\s$ contributed similarly to the curvature and isocurvature components
in the CMB, the result is compatible with current constraints.

\section{Phenomenology}

We now turn to the observational predictions of shift-symmetric orbital inflation. {For any positive $\mathcal{C}$, from} \eqref{spectrum}, the tensor-to-scalar ratio can be expressed as $r=16\epsilon_\ast/(1+\mathcal{C})$, and the scalar spectral index is $n_s-1 \equiv\frac{d \ln   P_\R}{d \ln k}=-2\epsilon_\ast-\eta_\ast+(d\mathcal{C}/dN)/(1+\mathcal{C})$,
where we used $d \ln k= dN$. Note that $\frac{\partial N_\ast}{\partial N} = -1$, since $N_\ast$ counts the number of efolds backwards. These predictions depend on the function $H(\theta)$. As in single field inflation, this function determines how slow-roll parameters $\epsilon$ and  $\eta \equiv {\epsilon^\prime}/{\epsilon}$ scale with $N_\ast$.

\begin{figure}[h]
\includegraphics[width=0.45\textwidth]{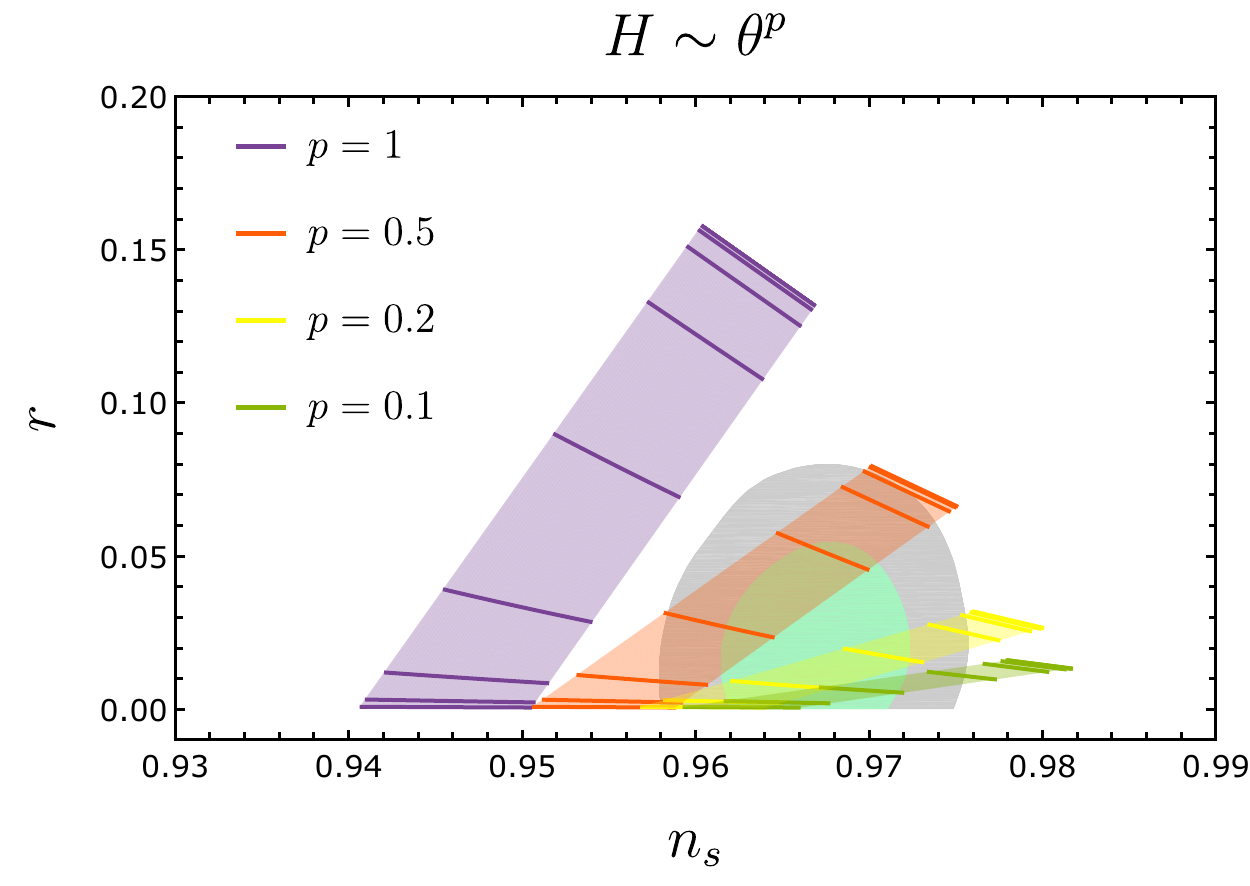}
\centering
\caption{The analytical predictions \eqref{eqn:nsrpowerlaw} for $(n_s, r)$ compared to the \textit{Planck}  $1\sigma$ and $2\sigma$ contours \cite{Akrami:2018odb}.
We show the predictions for wavenumbers which cross the horizon $50-60$ efolds before the end of inflation. The predictions for $n_s-r$ depend on the value of $\kappa \in [1,1000]$, where the values $(1, 2, 4, 8, 16, 32, 64, 128, 256)$
are depicted with thick lines (from bottom to top).}
\label{fig:nsrforpowerlaw}
\end{figure}

For concreteness, we consider models with $H\sim \theta^p$. {Solving \eqref{eqn:exactsolutionbgchapexactmodels} for $\theta(N)$} yields\footnote{{We note that for $0<p<1$ this toy model is not well defined as $\theta \to 0$, as can be seen in \eqref{potentials}. This is not a problem as the inflationary period we are interested in occurs before that point is reached. The true underlying potential would have to be completed in some way. This is similar to the case with say axion monodromy.}}
$\epsilon_\ast \simeq {p}/{(2N_\ast)}$ and $\eta_\ast \simeq {1}/{N_\ast}$.
The predictions for $n_s$ and $r$ are therefore well approximated by
\be\label{eqn:nsrpowerlaw}
n_s-1 \simeq - \frac{p+1}{N_\ast} - \frac{4p}{{\kappa^2}+4p N_\ast}, ~~ r\simeq \frac{8p \kappa^2}{N_\ast \kappa^2+4p  N_\ast^2} .
\ee
We plot these results against the Planck  $1\sigma$ and $2\sigma$ contours \cite{Akrami:2018odb} in Fig.~\ref{fig:nsrforpowerlaw}.
$N_\ast$ is taken to be between 50 and 60, and the radius $\kappa^2$ varies between $1$ and $10^5$.
The purple region is for $p=1$, corresponding to the toy model \eqref{eqn:toymodel}, and we also show the predictions for $p=0.5$ (red region), $p={0.2}$ (yellow region) and $p={0.1}$ (green region).

Notice that $n_s$ and $r$ only depend on the value of $\kappa$ and are therefore insensitive to the details of the field metric. When $\kappa\rightarrow\infty$ one recovers the predictions of chaotic inflation with $V\propto\phi^{2p}$. Meanwhile as $\kappa$ decreases, predictions are pushed downwards and to the left in this $n_s-r$ diagram. Therefore, in the case of power-law potentials only for small $p$ do the predictions remain within the {Planck} contours. The interesting regime here is still the case with significant turning (small $\kappa$ or $\mathcal{C}\gg1$), where the final power spectrum $P_\R \simeq \frac{H_*^2N_\ast^2}{\pi^2\kappa^2}$ mainly has an isocurvature origin. Then the tensor-to-scalar ratio is given by $r= 2\kappa^2/N_\ast^2=16\epsilon_\ast/\mathcal{C}$, which is suppressed. The spectral index reduces to $n_s-1=-(p+2)/N_\ast$ which, for small $p$, lies in the sweet spot $n_s =0.9649\pm0.0042$.

Another important observable is primordial non-Gaussianity, {which is currently bounded by Planck through $f^{\rm loc}_{\rm NL}=0.8\pm 5$~\cite{Ade:2015ava}}. There are examples in the literature of how $\mathcal{O}(1)$ local non-Gaussianity can arise in multi-field models, especially when the coupling between isocurvature and curvature modes is large \cite{Lyth:2005fi, Byrnes:2008wi, Byrnes:2009qy, Chen:2009zp} - see \cite{Byrnes:2010em} {for a review}. {There are also examples of how small levels of non-Gaussianity can arise in multifield models \cite{Dias:2017gva, Bjorkmo:2017nzd, Hotinli:2017vhx}. However, in most cases a detailed analytic understanding of the size of the non-Gaussianity is lacking because the associated dynamics is non-linear and complicated. This is not the case in  shift-symmetric orbital inflation, where we find that we can both easily satisfy the Planck constraint and crucially understand its origin analytically}. The amplitude of local non-Gaussianity can be determined using the $\delta N$ formalism. In a generic multi-field inflation model with curved field manifold,  we have $f^{\rm loc}_{\rm NL} =  \frac56 {G^{ab}G^{cd}N_aN_cN_{bd}}/{(G^{ab}N_aN_b)^2}$ \cite{Lyth:2005fi, Seery:2005gb}, where $G_{ab}={\rm diag}\{f(\rho), 1\}$ is the field space metric, $N_a$ and $N_{ab}$ are  derivatives of $N$ with respect to the fields $(\theta, \rho)$.
To gain some analytical understanding, here we still focus on models with $H\sim \theta^p$, where $N$ can be expressed as $N=f(\rho)\theta^2/4p-p/2$.
The amplitude of local non-Gaussianity then follows
\be \label{fnl}
 f^{\rm loc}_{\rm NL}
=\frac{5}{12}\eta_\ast\left[1-\frac{\mathcal{C}^2}{(1+\mathcal{C})^2}\frac{\kappa^2\mathbb{R}}{2}\right],
\ee
where we used the relation $\mathcal{C}=2p^2/(\epsilon_\ast\kappa^2)$. 
When $\kappa\rightarrow\infty$, we have $\mathcal{C}\rightarrow0$ and $\mathcal{C}^2\kappa^2\rightarrow0$.
Thus the second term in \eqref{fnl} vanishes, which leads to the single field result $f^{\rm loc}_{\rm NL} ={5\eta_\ast}/{12}$ as expected. The enhancement of non-Gaussianity is possible in the intermediate regime $\mathcal{C}\sim \mathcal{O}(1)$, where the transfer from isocurvature to adiabatic modes is inefficient. In that case, $f^{\rm loc}_{\rm NL} \sim -5p \mathbb{R}/12$ can be large if the field space is highly curved.

For the interesting regime with $\mathcal{C}\gg1$, the $\delta N$ expansion is dominated by $N_\rho$ and $N_{\rho\rho}$. This then leads to what, at first sight, appears as the counterintuitive result that $f^{\rm loc}_{\rm NL}$  is negligible and slow-roll suppressed
\begin{equation} \label{fnl-C}
 f^{\rm loc}_{\rm NL} \simeq \frac{5}{6}\frac{N_{\rho\rho}}{N_\rho^2} = \frac{5}{12}\eta_\ast\left(1-\frac{\kappa^2 \mathbb{R}}{2}\right) .
\end{equation}
This is {the same as} happened in the calculation of {the} power spectrum: the contribution to the curvature perturbation sourced by the isocurvature modes dominates the final result. {The bispectrum is found to be slow-roll suppressed, just like in single field inflation, but there are small corrections from {the} field space curvature, which violates Maldacena's  consistency relation \cite{Maldacena:2002vr, Creminelli:2004yq}. We have recently confirmed this result via a scaling symmetry approach {in}~\cite{Achucarro:2019lgo}}.

\section{Discussions}

We have proposed  a class of multi-field inflationary models that demonstrate a new type of attractor trajectory along the isometry direction in field space. Here the isocurvature modes become massless and freeze on superhorizon scales.  Moreover, when the turning effects become significant, the curvature perturbations keep growing after horizon-exit and thus isocurvature modes  are dynamically suppressed.  As a consequence, these multi-field models yield  the single-field-like phenomenology favored by observations.

Additional isocurvature perturbations 
will either decay if they are massive or 
freeze if they are light.  Therefore, although our computations were done in a simple two-field setting, we expect the conclusions will continue to hold in multi-field extensions with more than two fields, provided that the number of additional light isocurvature fields is not too large.

We have shown and explained how in shift-symmetric orbital inflation, a negligible amount of local non-Gaussianity is produced.
Here the isocurvature degree of freedom can be the dominant contribution to the bispectrum, but in such cases $f_{\rm NL}$ is slow-roll suppressed. This result teaches us a generic lesson: that in multi-field models, even if the isocurvature-to-adiabatic conversion is very efficient, the resulting non-Gaussianity can still be suppressed.
A large coupling between curvature and isocurvature modes enhances the transfer of non-Gaussianity, but for this transfer to generate large non-Gaussianity, one needs sizable self-interactions affecting the isocurvature field during horizon crossing~\cite{Chen:2009zp, Chen:2018uul}. {In this class of scenarios, however, the shift symmetry along the radial direction (\ref{symmetry}) has a role in suppressing the self-interactions of the isocurvature field (see \cite{Achucarro:2019lgo}).}
Therefore, it is perfectly fine to study multi-field models with significant and sustained turning trajectories, without worrying about generating large non-Gaussianity.

Our model has important implications on the realization of inflation in UV-complete theories. Contrary to what is usually assumed, and as emphasized in~\cite{Achucarro:2016fby}, it is not always necessary to stabilize all compactification moduli, or to have a large mass hierarchy between the inflaton and other fields. 
The suppression of isocurvature perturbations and non-Gaussianity has the common origin in shift-symmetric orbital inflation. 
From an EFT point of view this can be traced back to the effect of derivative interactions among the curvature and isocurvature perturbations that are absent in single-field inflation. These are unavoidable on curved trajectories and curved field spaces and, therefore, ubiquitous in string compactifications.

{\it Acknowledgments.--}
We are grateful to Valeri Vardanyan
for collaboration in the early stages of this work.
Also to Martina Chirilus-Bruckner, Cristiano Germani and Gabriel Jung for stimulating discussions and
comments. We are very grateful to the referees for their constructive and helpful comments. The work of AA is partially supported by the Netherlands' Organization for Fundamental Research in Matter (FOM), by the Basque Government (IT-979-16) and by the Spanish Ministry MINECO (FPA2015-64041-C2-1P).
EJC is supported by STFC Consolidated Grant No. ST/P000703/1 and would like to acknowledge the wonderful support and hospitality of colleagues at the Lorentz Institute where some of this work was carried out. GAP acknowledges support from the Fondecyt Regular Project No. 1171811 (CONICYT).
OI, DGW and YW are supported by the Netherlands Organization for Scientific Research (NWO) and OCW.
YW is also supported by the ERC Consolidator Grant STRINGFLATION under the HORIZON 2020 grant agreement no. 647995.

\section*{Appendix A. Hamilton-Jacobi Formalism}
\renewcommand{\theequation}{A\arabic{equation}}
\setcounter{equation}{0}
Here we apply the Hamilton-Jacobi formalism \cite{Muslimov:1990be, Salopek:1990jq,Lidsey:1991zp,Copeland:1993jj} to derive the potential for shift-symmetric orbital inflation, {replacing the potential as an input function with the Hubble parameter $H(\phi)$  which leads directly to the inflation dynamics}.

Friedmann's {second} equation yields
\begin{equation} \label{eq:HJ2nd}
 \dot H = \dot\phi H_\phi  = -\frac{\dot\phi^2}{2} \, \longrightarrow \,  -2H_\phi=\dot\phi \, .
\end{equation}
{leading to the Hamilton-Jacobi form of the first Friedmann equation}
\begin{equation}
V=3H^2-2H_\phi^2 \, , \label{eq:HamiltonJacobisingle}
\end{equation}
with all functions now being explicitly dependent on $\phi$. Therefore, if $H(\phi)$ is known, {so is $V(\phi)$}.

For the multi-field case, {we simply generalise, $\phi \to \phi^a$, so equations (\ref{eq:HJ2nd}) and (\ref{eq:HamiltonJacobisingle}) become}
\begin{equation}
 \dot H = \dot\phi^a H_a  = -\frac{\dot\phi^a\dot\phi^b G_{ab}}{2} \, \longrightarrow \,  H_a = -\frac{G_{ab}\dot\phi^b}{2} \ .
 \label{eqn:HJdynamicstwofield}
\end{equation}
{ and}
\begin{equation}
3H^2 = V + 2  H^a H_a .
\label{eqn:HJtwofield}
\end{equation}
{ which we use} to construct the generic potentials for shift-symmetric orbital inflation.
The important requirement here is that {the} inflaton trajectory is along the isometry direction at any radius.
Thus for the field space $(\theta, \rho)$ with metric $G_{ab}={\rm diag}\{f(\rho), 1\}$,
{the} inflaton should move in the $\theta$ direction for any value of $\rho$.
For this behaviour, equation \eqref{eqn:HJtwofield} simplifies to
$ 3H^2  = V + 2  \frac{H_\theta^2}{f(\rho)}$.
Therefore, we conclude that our two-field inflationary model has a potential of the following form
\begin{equation}
V= 3H(\theta)^2  - 2  \frac{H_\theta^2}{f(\rho)} .
\label{eqn:familyofpotentials}
\end{equation}

\section*{Appendix B. Stability Analysis}
\label{sec:stability}
\renewcommand{\theequation}{B\arabic{equation}}
\setcounter{equation}{0}
Here we demonstrate the neutral stability of the exact solutions. We have seen that there is a continuous set of orbital solutions parametrized by $\rho_0$ and that normal perturbations move us freely between these `attractors', so the system is not stable in the usual sense. The property we need to prove is that small perturbations shift us to another inflationary solution $\dot\rho = 0$.

Each attractor solution
$$\dot\theta  = -2 \frac{H_\theta}{f}, ~~~~ \rho = \rho_0$$ corresponds to a point in the $(\dot\rho, \dot\theta)$ plane. These points are all different and lie on a curve, therefore the stability of this system is non-trivial to prove analytically. If we simply perturb the field equations we will find zero eigenvalues associated with the perturbations that move us between attractors. Moreover, it is not obvious how to find variables such that the linearized system of perturbations becomes diagonal.  Our approach is to introduce the variables
\ba
x(\theta,\rho, \theta^\prime, \rho^\prime) &\equiv \frac{f H}{ H_\theta}\theta^\prime - 2\frac{f}{f_\rho}{\rho^\prime} +2, \\
y(\theta,\rho,\theta^\prime, \rho^\prime)  &\equiv \frac{f H}{ H_\theta}\theta^\prime +2\ ,\\
 z(\theta,\rho) &\equiv \frac{f H^2}{ H_\theta^2}-2/3\, ,
\ea
here a prime denotes a derivative with respect to {the number of} efolds $(..)^\prime = \frac{d}{dN}(..)$. Remember that $H=H(\theta)$ and $f=f(\rho)$.
Our definition of stability now amounts to the presence of a fixed point at $(x, y) = (0,0)$.

For $H\sim \theta$ the potential in \eqref{eqn:familyofpotentials} satisfies the following scaling relation
\begin{equation}
 \theta V_\theta - 2\frac{f}{f_\rho} V_\rho = 2 V.
 \label{eqn:scalingrelationpotential}
\end{equation}
This ensures that the equations for $x$ and $y$ diagonalize at the linear level, {and below we prove linear stability for the models $H\sim \theta$, although it applies to any power law $H\sim \theta^n$ and more general models.}

\subsection{Linear stability analysis}\label{sec:stabilitygeneralH}
{In terms of $x$, $y$, $\rho$ and $z$, the field equations and second Friedmann equation become}
\ba
&x^\prime + (3-\epsilon)x +\left(2\left(\frac{f}{f_\rho}\right)_\rho-g(\theta)\right)\left({\rho^\prime}\right)^2 \\
&+\frac{2(z+2/3)}{z}g(\theta)\left(\epsilon - \epsilon_0\right) = 0, \notag\\
&y^\prime + (3-\epsilon)y  +\frac{2}{z}\left(-\frac{1}{3}\left({\rho^\prime}\right)^2-\frac{1}{2}y^2+2y\right) \\
&-g(\theta)\left({\rho^\prime}\right)^2  +\frac{2(z+2/3)}{z}g(\theta)\left(\epsilon - \epsilon_0 \right) = 0 , \notag  \\
&z^\prime = 2(y-2)\left(1-g(\theta) \right)+ \left(\frac{ f_\rho}{f}\right)^2\frac{y-x}{2}\left(z+\frac{2}{3}\right),\\
&{\rho^\prime} = \frac{ f_\rho}{f}\frac{y-x}{2},\\
&\epsilon = \frac{1}{2}\frac{(y-2)^2}{z+2/3}+\frac{ f_\rho^2}{f^2}\frac{(x-y)^2}{8},
\ea
where $\epsilon_0 = \frac{2}{z+2/3}$. All the terms in brackets are combined to be manifestly zero on the {attractor, and we have} introduced the model specific function $g(\theta) \equiv \frac{H H_{\theta\theta}}{H_\theta^2}$. Note that $g(\theta)$ is in general a function of $z$ and $\rho$,  but it reduces to a constant in the case when we have a power law $H(\theta) \sim \theta^n$, and it is zero for $n=1$.

In terms of the four variables, shift-symmetric Orbital Inflation is given by $(x,y,z^\prime,\rho^\prime) = (0,0,-4(1-g(\theta)),0)$. {To prove it is the attractor we must show $(y,\rho^\prime)=(0,0)$ is a fixed point}.
Note that the friction term is very large during inflation. We can already see that without the friction the system would be {unstable, so we now establish whether the friction term is in fact} large enough to make the system stable.

{ Linearly perturbing around $(y,\rho^\prime) = (0,0)$ with $\epsilon = \frac{2}{z+2/3}$ yields}
\ba
&\delta x^\prime + \left(3- \frac{2}{z+2/3}\right) \delta x - \frac{4 g(\theta)}{z}\delta y = 0, \label{eqn:stabilitydeltax2} \\
&\delta y^\prime + \left(3- \frac{2}{z+2/3}+\frac{4(1-g(\theta))}{z}\right) \delta y = 0, \label{eqn:stabilitydeltay2} \\
&\delta z^\prime = 2(1-g(\theta))\delta y +  \left(\frac{ f_\rho}{f}\right)^2\frac{\delta y-\delta x}{2}\left(z+\frac{2}{3}\right), \\
&{\delta \rho^\prime} =  \frac{ f_\rho}{f}\frac{\delta y-\delta x}{2}. \label{eqn:stabilitydeltarho2}
\ea
For constant $g(\theta)$ {below} we  explicitly prove stability. For a  general  $g(\theta)$ {we express it} in terms of $z$ and $\rho$ and integrate the equations numerically. However, we expect the system to be stable. If $(1-g(\theta))$ takes values of order 1 and does not vary too rapidly, then $z$ will take large values during inflation and behave smoothly as well. In that case we see from \eqref{eqn:stabilitydeltax2} and \eqref{eqn:stabilitydeltay2} that $\delta x^\prime$ and $\delta y^\prime$ are dominated by the friction terms $-3\delta x$ and $-3\delta y$ respectively. Therefore, we expect both of them to decay like $e^{-3N}$. Finally \eqref{eqn:stabilitydeltarho2} then implies that we quickly converge to the fixed point.

\subsubsection{Power law inflation $H\sim \theta^n$}\label{sec:stabilitypowerlawH}
In the case of power law inflation with $1- g(\theta) = \frac{1}{n}$, { using $z=z_0 - \frac{4}{n}N$, we can solve \eqref{eqn:stabilitydeltay2} and \eqref{eqn:stabilitydeltax2} yielding}
\ba
 &\delta x = \delta x_0 \left(\frac{2+3 z_0 }{2+3 z} \right)^{n/2}e^{-3N}
  +\delta y_0 \frac{4(n-1) N}{n}\left(\frac{2+3 z_0 }{2+3 z} \right)^{n/2}e^{-3N} ,\notag \\
   &\delta y  = \delta y_0 \frac{z}{z_0}\left(\frac{2+3 z_0 }{2+3 z} \right)^{n/2}e^{-3N},
\label{eqn:stabilitydeltarho}
\ea
{ which in \eqref{eqn:stabilitydeltarho2} demonstrates}
that $(y,\rho^\prime) = (0,0)$ is a fixed point. This proves stability for power law inflation.

\subsubsection{Linearized equations in {the} slow-roll parameters}
We can write the linearized perturbation equations in terms of the slow-roll parameters $\epsilon$ and $\eta$
\begin{equation}
\epsilon = \frac{2 H_\theta^2}{fH^2}, \quad \eta \equiv \frac{\dot\epsilon}{H \epsilon}= -\frac{4 H_{\theta\theta}}{fH} + \frac{4}{f}\left(\frac{H_\theta}{H}\right)^2\ .
\end{equation}
In particular, the model specific function $g(\theta)$ becomes
\begin{equation}
g(\theta) = (2\epsilon-\eta)\sqrt{\frac{f}{8\epsilon }}\ .
\end{equation}
We see that $g(\theta)$ is not necessarily positive, but it will be small if both the slow-roll approximation and the condition $\eta \ll \sqrt{\epsilon}$ hold true.
The linearized equations \eqref{eqn:stabilitydeltax2} -- \eqref{eqn:stabilitydeltarho2}  are then given by
\ba
&\delta x^\prime + \left(3- \epsilon \right) \delta x - \frac{2\epsilon-\eta}{1-\epsilon/3}\sqrt{\frac{\epsilon f}{2 }}\delta y = 0, \\
&\delta y^\prime + \left(3- \epsilon+\frac{1}{1-\epsilon/3}\left(2\epsilon-(2\epsilon-\eta)\sqrt{\frac{\epsilon f}{2 }}\right)\right) \delta y = 0, \nonumber\\
&\delta z^\prime = 2\left(1- (2\epsilon-\eta)\sqrt{\frac{f}{8\epsilon }}\right)\delta y +  \left(\frac{ f_\rho}{f}\right)^2\frac{\delta y-\delta x}{\epsilon}, \nonumber \\
&{\delta \rho^\prime} =  \frac{ f_\rho}{f}\frac{\delta y-\delta x}{2}.\nonumber
\ea
In the slow-roll approximation $\delta x$ and $\delta y$ are therefore exponentially decaying
\ba
&\delta x \approx \delta x_0 e^{-3N}, \\
& \delta y \approx \delta y_0 e^{-3N}.
\ea
Looking at the equation for $\delta z$ we find that a sufficient condition for stability is that $e^{-3N}/\epsilon$ goes to zero exponentially fast. This requires $\eta < 3$, which is automatically satisfied assuming the slow-roll approximation $\eta \ll 1$.  In addition, $\epsilon$ cannot be arbitrarily small.

\bibliography{bibfile}

\end{document}